\documentclass[preprint,sort&compress,5p,twocolumn]{elsarticle}
\usepackage[british]{babel}
\usepackage{amsmath,amstext,graphicx}
\usepackage[utf8]{inputenc}
\usepackage[T1]{fontenc}
\usepackage{mathptmx}
\usepackage{color}
\usepackage{lineno}
\usepackage{tabularx}
\usepackage[unicode,pdfborder={0 0 0}, colorlinks=false]{hyperref}

\journal{Nucl.~Instr.~Methods Phys.~Res., Sect.~A}

\begin{document}
\setlength\linenumbersep{5pt}
%
\begin{frontmatter}
\title{Single-particle detection of products from atomic and molecular reactions\\in a cryogenic ion storage ring}
\author[MPI]{C.~Krantz\corref{cor1}}
\author[MPI]{O.~Novotný\corref{cor2}}
\author[MPI]{A.~Becker}
\author[MPI]{S.~George}
\author[MPI]{M.~Grieser}
\author[MPI]{R.~von~Hahn}
\author[MPI]{C.~Meyer}
\author[UniGI]{S.~Schippers}
\author[MPI,IAMP]{K.~Spruck}
\author[MPI]{S.~Vogel}
\author[MPI]{and A.~Wolf}
\cortext[cor1]{claude.krantz@med.uni-heidelberg.de --- present address: Marburg Ion-Beam Therapy Centre, 35043 Marburg, Germany}
\cortext[cor2]{oldrich.novotny@mpi-hd.mpg.de}

\address[MPI]{Max-Planck-Institut für Kernphysik, Saupfercheckweg 1, 69117 Heidelberg, Germany}
\address[UniGI]{I.\ Physkalisches Institut, Abt.\ Atom- und Molekülphysik, Justus-Liebig-Universität Gießen, Heinrich-Buff-Ring 16, 35390, Gießen, Germany}
\address[IAMP]{Institut für Atom- und Molekülphysik, Justus-Liebig-Universität Gießen, Leihgesterner Weg 217, 35392 Gießen, Germany}
\begin{abstract}
  We have used a single-particle detector system, based on secondary electron emission, for counting low-energetic ($\sim$ keV/u) massive products originating from atomic and molecular ion reactions in the electrostatic Cryogenic Storage Ring (CSR). The detector is movable within the cryogenic vacuum chamber of CSR, and was used to measure production rates of a variety of charged and neutral daughter particles. In operation at a temperature of $\sim 6$~K, the detector is characterised by a high dynamic range, combining a low dark event rate with good high-rate particle counting capability. On-line measurement of the pulse height distributions proved to be an important monitor of the detector response at low temperature. Statistical pulse-height analysis allows to infer the particle detection efficiency of the detector, which has been found to be close to unity also in cryogenic operation at 6~K.   
\end{abstract}
\end{frontmatter}
\section{Introduction}
Single-particle counting detectors are important instruments in many atomic and molecular physics experiments on fast-propagating ion beams \cite{WolfDR,SchippersAto}. In such experiments, an ion beam is guided through a target medium which can consist, e.g., of photons, electrons, neutral atoms, or molecules. Reactions of the projectile ions with the target particles typically lead to products of different charge-to-mass ratio. This results in the formation of daughter beams of different ion-optical rigidity compared to the parent, which can be separated from the latter by electric or magnetic analysing fields. At known intensity of the parent beam and thickness of the target, detection of the daughter particles reveals the rate coefficients of the processes involved in their production. Due to the typically low ion numbers and reaction cross-sections, the product detection needs to be done on the single-particle level.\par
Heavy-ion storage rings enhance such target experiments by their ability to store the projectiles for extended periods of time. Due to energetic processes in the ion source, unknown, highly-excited quantum states are often populated in atomic or molecular ions directly after production. In may cases storage of the ions enables them to reach a well-understood state population by spontaneous decay before undergoing the actual experiment. The extended storage time also allows phase-space manipulation of the ion beam, such as electron or stochastic cooling, or initial-state preparation techniques as required for laser- or collision-driven pump-probe experiments \cite{pedersen,stoechkel}.\par 
For years, medium-energy magnetic ion synchrotrons have been used very successfully for these kinds of experiments---a remarkable development considering that the technology of those machines was originally aimed at nuclear physics applications \cite{astrid,cryring,TSRIsolde}. Based on that success, a new class of heavy-ion storage rings has emerged, with designs that are optimised for experiments on atomic and molecular physics. They use purely electrostatic ion optics, matching the output energy of relatively simple electrostatic injectors that can be flexibly equipped with state-of-the art molecular ion sources \cite{elisa,kek,schmidtessr}. The most advanced set-ups use cryogenic cooling machines to reduce the temperature of their beam guiding vacuum vessels down to values near that of liquid helium \cite{tmu,desiree,thomasdesiree,riken,csrpaper}. On the one hand, this results in a vastly improved residual gas pressure compared to conventional ultra-high vacuum (UHV) set-ups, with correspondingly longer ion storage times \cite{langectf,backstroemPRL}. On the other hand, storage in such a cold environment allows infra-red-active molecular ions to de-excite to their lowest rovibrational levels prior to starting experiments---a significant improvement over room-temperature ion-storage facilities \cite{ZajfmanDR,oconnorCH}.\par
The advantages of these cryogenic ion storage rings come with technological challenges with respect to the particle detector equipment. A restriction regarding possible detection principles arises from the low energy of the product particles. Limited by available high-voltage technology, typical kinetic energies in electrostatic storage devices are of order a few keV/u or below. This rules out detection mechanisms where the counting volume of the detector is covered by significant layers of passive material---as is the case for surface-barrier semi-conductor counters \cite{sheehan} and, to lesser extent, for scintillators \cite{akthar}. Open micro-calorimetric detectors are a promising option for product detection at cryogenic storage rings, which is presently under investigation \cite{novotnyMMC,gamerMOCCA,ohkubo}. Their fabrication and operation are however extremely difficult and expensive, such that their use may be limited to selected experiments in the foreseeable future.\par 
Suitable detectors for cryogenic storage rings, which can be widely deployed at acceptable manufacturing and operating costs, are therefore based on surface secondary-electron emission with subsequent multiplication \cite{thomasdesiree,csrpaper}. This detection technique has proven itself also at particle energies below 1~keV/u \cite{rinn1982}, but the low-temperature environment does come with new challenges. Besides engineering problems related to thermal expansion and embrittlement of materials, the efficiency of charge multiplication stages commonly used in low-energy ion detection is known to suffer in cold operation. Due to their semi-conductor-like properties, the electric resistance of micro-channel plates (MCPs) and single-channel electron multipliers (CEMs) rises strongly upon cooling into the cryogenic regime. The high resistance can lead to decreased gain or even complete charge depletion, especially at elevated particle hit rates. Depending on the application, MCPs have been used near $\sim 10$~K with varying degrees of success \cite{schecker,roth,kuehnel}. Even less is known about the low-temperature behaviour of CEMs \cite{sawicki}.\par 
In a recent publication, we have presented the design of a movable single-particle counting detector for the Cryogenic Storage Ring (CSR) of the Max Planck Institute for Nuclear Physics (MPIK) in Heidelberg, Germany \cite{spruck}. Here, we report on the first operation of this device under real-life experimental conditions at the CSR.\par
This paper is structured as follows: In Section~\ref{experimental} we briefly describe the instrument. In Section~\ref{operation} we present the most important findings from the first operation of the detection system with the storage ring CSR at its lowest temperature of $\sim 6$~K. In Section~\ref{results} we quantify and discuss the results from that series of experiments, with emphasis on the single-particle detection efficiency of the set-up. Section~\ref{summary} closes with a summary and outlook onto future developments.  
\section{Overview of the Experimental Set-Up\label{experimental}}
The CSR is a fully electrostatic storage ring designed for positive or negative ions of kinetic energies up to 300~keV per unit of charge  \cite{csrpaper}. The beam guiding vacuum vessel as well as the ion optics contained therein can be cooled to temperatures of $\sim$6~K by a closed-loop liquid-helium refrigerator. For thermal insulation, the beam line is enclosed in an additional isolation vacuum vessel and protected by several layers of black-body-radiation shields.\par
\begin{figure}[tb]
  \centering
  \includegraphics[width=\columnwidth]{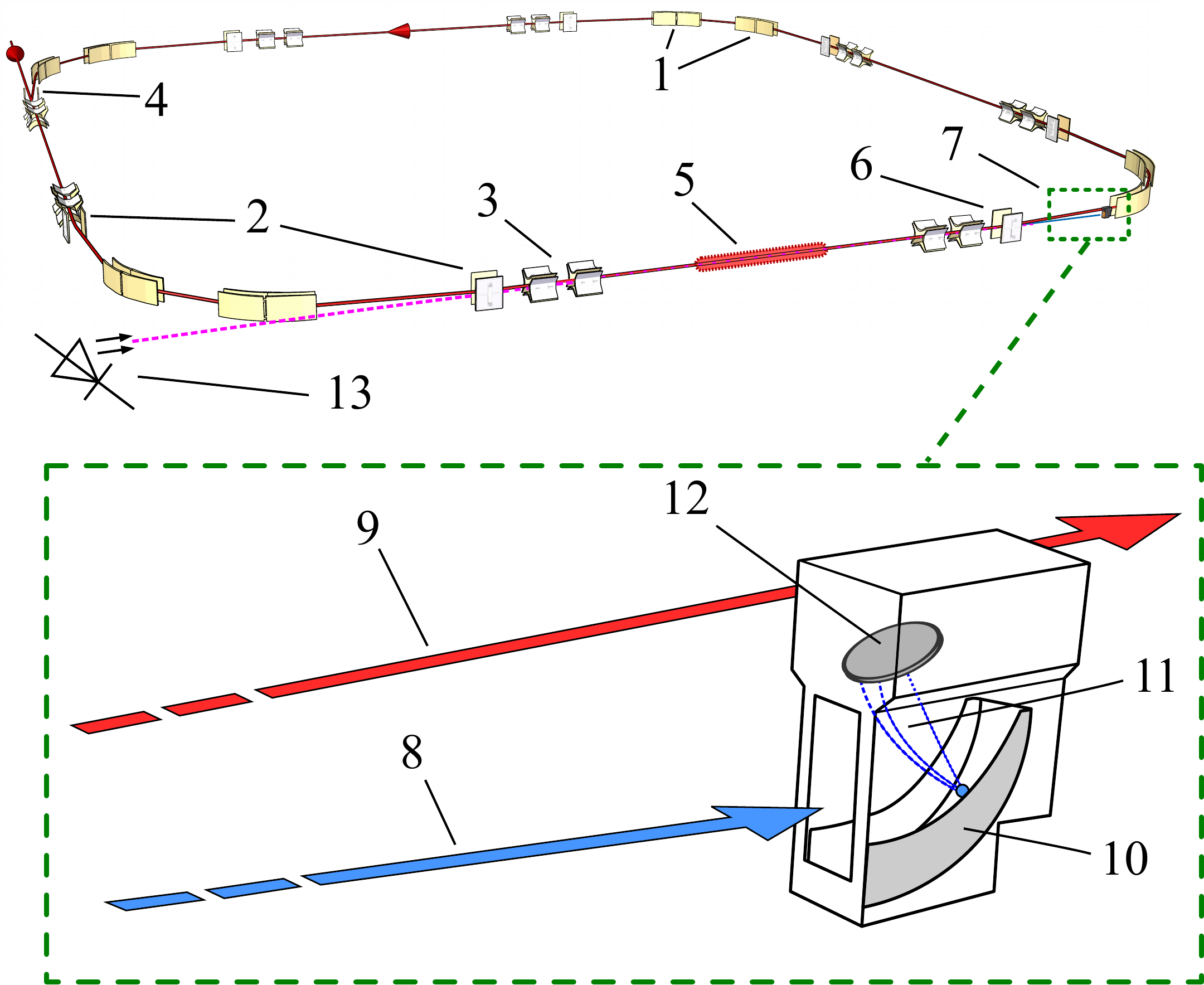}
  \caption{Schematic view of the experimental set-up, consisting of the storage ring CSR \cite{csrpaper} (top) and the COMPACT detector \cite{spruck} (bottom). The lattice of CSR is a four-fold symmetric, approximately quadratic, and purely electrostatic beam-line, consisting of a total of eight 39$^\circ$ deflectors (1), eight 6$^\circ$ deflectors (2), and eight focussing quadrupole doublets (3). One of the 6$^\circ$ deflectors is fast-switchable to allow  ion beam injection (4). In many of the here-reported measurements a laser beam overlapped the stored ions in an experimental target section (5). The 6$^\circ$ deflector (6) directly following the target acts as charge-to-mass analyser preceding the movable COMPACT detector (7). The latter can be positioned to intercept the product particles (8) from atomic reactions in the target while allowing the parent ion beam (9) to circulate unhindered in the CSR. In the detector, the product particles hit a secondary-electron emitting cathode (10). These electrons (11) are accelerated towards a micro-channel plate stack (12) where they are multiplied to form the detector current pulse. For off-beam testing, the detector can be irradiated by an ultra-violet (UV) light emitting diode (LED, 13) installed in the opposite sector of CSR. For details see text and Refs.\ \cite{csrpaper, spruck}.\label{csrfigure}}
\end{figure}

 
With an orbit circumference of 35~m, the storage ring (cf.\ Fig.~\ref{csrfigure}) consists of four identical ion-optical sectors which enclose four field-free drift sections. While one of the latter is occupied by the beam diagnostic instrumentation of the storage ring \cite{grieseripac}, the other three are free for installation of experimental equipment. The counting detector (lower panel of Fig.~\ref{csrfigure}) is located downstream from an experimental section, within one of the ion-optics sectors of CSR. The technology of the detector system has been described extensively in a dedicated publication \cite{spruck}, hence we limit ourselves to a brief overview here.\par 

Equipped with a 20-mm-wide entrance window for heavy particles, the detector is movable transversely to the beam direction in the plane of the storage ring. It is installed 1.0~m downstream of a short (6$^\circ$) electrostatic bending dipole of the storage ring. Product particles generated from the stored ions are deflected at a characteristic angle in the dipole element. By placement at a suitable horizontal position, the detector can intercept products with a charge-to-mass ratio that differs from that of the stored parent beam by more than 100\,\% in both directions. Specifically, it can detect neutral products on axis of the ion beam in the experiment as well as, e.g., ionisation products up to the double charge of a stored atomic cation beam \cite{spruck}.\par
Eventually, the detector is designed to intercept product particles originating from ion-electron interactions in the future electron cooler of CSR---like electron recombination or electron impact ionisation \cite{KrantzDR,HahnIonis}. In contrast to the detector set-up, the cooler was not yet operational during the 2015 experiments. Instead, an ion-photon interaction beam line was installed in the experimental CSR section preceding the detector \cite{csrpaper}. It allowed to overlap the stored ions at grazing angle with laser beams of various wavelengths that were coupled into CSR using a system of broadband view-ports and mirrors in the cryogenic vacuum chamber. This in-ring laser target was used in experiments on photo-induced electron detachment of stored anions. In addition, without using the laser beams, experiments on auto-detachment and auto-fragmentation of excited molecular and cluster ions were performed using the same set-up. At higher CSR operating temperatures, products of electron transfer from the residual gas to stored cations were observed. For testing purposes, the detector can be irradiated by photons from an ultra-violet (UV, 245(5)~nm) light emitting diode (LED) \cite{spruck}. The UV-LED is located in a room temperature annex of the CSR sector opposite of the detector. The beam of photons from the LED is practically uncollimated and enters the CSR vacuum chamber via a set of UV-grade sapphire view-ports.\par       
The detector employs a variant of the `Daly' ion detection principle, where incident massive particles impinge onto a sec\-ond\-ary-electron emitting cathode made of aluminium \cite{spruck,rinn1982,daly}. The secondary electrons released in each hit are accelerated by 1.2~kV towards a small chevron micro-channel plate stack (cf.\ Fig.~\ref{csrfigure}). The latter acts as secondary-electron multiplier, while being protected from direct hits by the primary massive ions. The multiplied electron bunches are then collected on a metal anode.\par 
After capacitive decoupling from high-voltage, the pulses are driven into a fast front-end amplifier of 50~$\Omega$ input imped\-ance and gain factor 200 (\emph{Ortec VT120A}). In most of the presented experiments, the resulting $\sim$\,10 ns-short electric pulses were converted into logical signals using a linear discriminator, and counted by a VME-based multiscaler. Simultaneously---but asynchronously---sample pulses could be recorded using a digital oscilloscope which served as waveform digitiser. This simple solution yields two independent datasets for the detector count rate and for the sample waveforms, which cannot be correlated on the single-particle level. In the course of the experiments, a second, more advanced data acquisition system was set up, consisting of a fast analog-to-digital converter (FADC, \emph{Agilent Acqiris U1084A}) equipped with a large sampling memory. This system allows gapless recording of the pre-amplified detector signal. Via an on-line peak-finding routine, it yields a single, consistent dataset containing the amplitude and arrival time of each individual detector pulse.\par
Much care was taken in preparing the detector to perform at cryogenic temperatures. With reference to that purpose, the device has been called `COMPACT', the `COld Movable PArticle CounTer' \cite{spruck}. In order to support optional room-temperature operation of CSR, the design additionally needed to fulfil the standard low-out-gassing requirements of bakeable UHV equipment. All electronics is kept on the atmosphere side of CSR's nested vacuum system, as is the rotary actuator that allows horizontal positioning of the particle sensor via a thread drive inside the CSR beam line.\par 
The chevron MCP stack consists of two matched, circular `extended dynamic range (EDR)' micro-channel plates (\emph{Photonis} 18/12/10/12 D 40:1 EDR, MS) of 18 mm useful diameter. EDR MCPs are characterised by a significantly lower resistance as compared to standard variants and are thus expected to perform better at very low detector temperatures. As an option to warm up the electron multiplier in operation, a small electric heater made of a bare Constantan wire is included in the supporting frame of the MCP stack \cite{spruck}.\par  
\section{Low Temperature Operation\label{operation}}
The first experimental beam-times at the CSR took place in 2015, and lasted for approximately five months, including cool-down of the storage ring by the liquid-helium refrigerator and rewarming of the facility \cite{csrpaper}. Besides the afore mentioned measurements on electron detachment and cluster fragmentation, a multitude of experiments with positive and negative ions were conducted in an effort to characterise the storage ring and beam diagnostic instrumentation. During most of the experiments, the CSR operated at an average temperature of $\sim$\,6~K. Due to technical issues of the injection accelerator, the ion energies were limited to $80$~keV, i.e.\ well below the CSR design energy of 300 keV per unit of charge. The stored ion species included Ar$^+$, N$_2^+$, O$^-$, OH$^-$, Si$^-$, C$_2^-$, Co$_2^-$, Co$_3^-$, and Ag$_2^-$ \cite{csrpaper}. The COMPACT detector system was used in almost all of the experiments, so that its low-temperature performance could be studied in a variety of use cases. This section presents a few examples of measurements that showcase the most important findings made during operation of the particle detector.\par  
\begin{figure}[tb]
  \centering
 \includegraphics[width=\columnwidth]{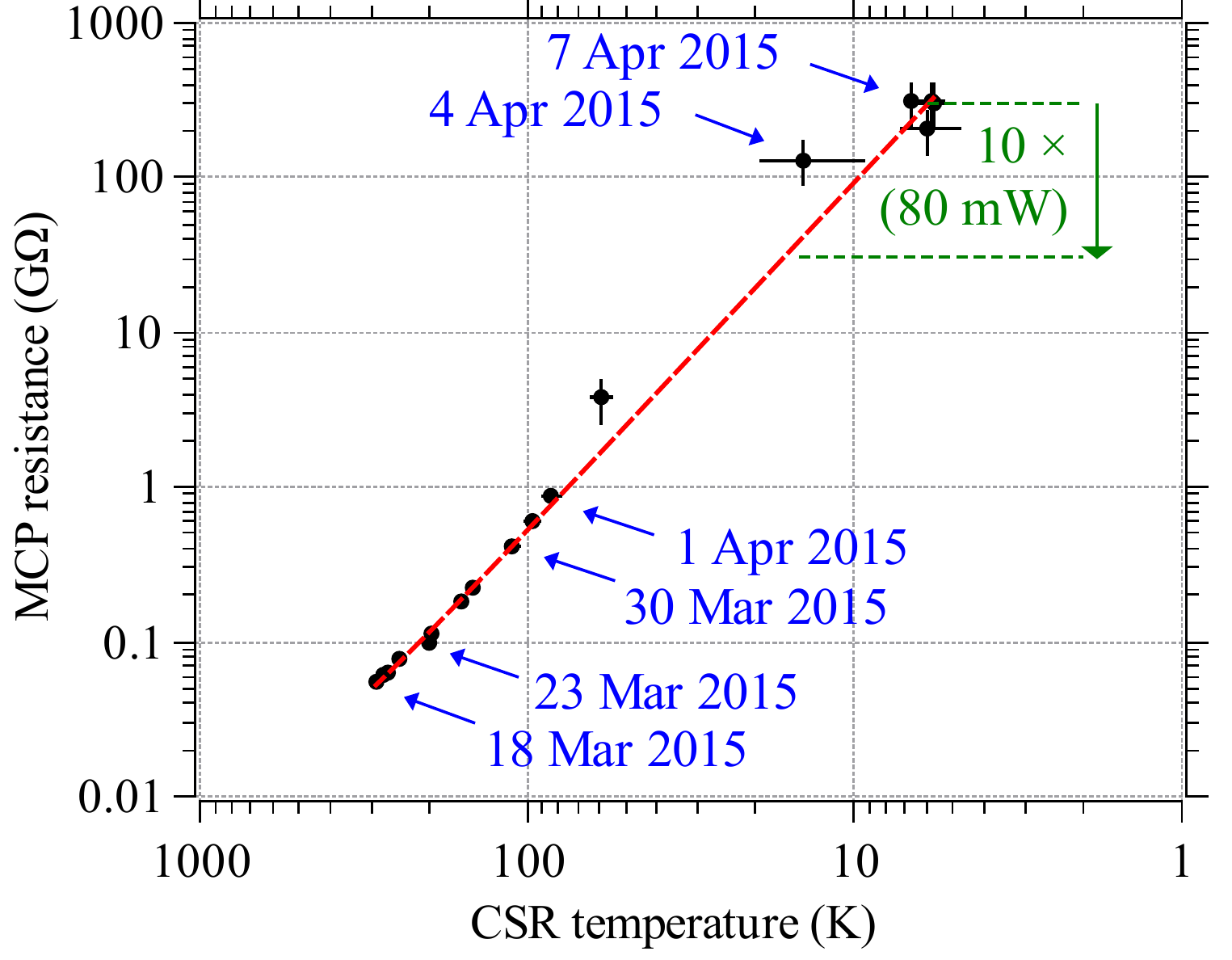}
 \caption{Electric resistance of the MCP set as a function of the ambient temperature (dots). The data was taken during the 2015 cool-down of the storage ring CSR from room temperature to $\sim$\,6~K, as indicated by the time stamps. The dashed line is a power-law fit to the data and intended as a guide to the eye only. The vertical arrow indicates the range within which the MCP resistance could optionally be varied using its built-in heater at a power of $\sim 80$~mW. In the reported single-particle experiments, the heater was not used and the detector was left at the 6\,(1)~K temperature of the surrounding CSR structures (cf. Sect.\ \ref{heating}).\label{mcptemp}}
\end{figure}
\subsection{Cool-Down}
While cryo-adsorption in cold operation vastly improves the residual gas pressure, the CSR vacuum concept does not rely on cryogenics alone. Before start of the cool-down procedure, the beam-guiding vacuum vessel of the storage ring was UHV-baked at 180$^\circ$C and subsequently reached a residual gas pressure of $\sim 1\times 10^{-10}$~mbar already at room temperature \cite{csrpaper}. Consequently, the storage ring and the detector could already operate during the cool-down phase of CSR from 300~K to 6~K, which took approximately three weeks.\par 
During the cool-down, a beam of 60-keV (1.5~keV/u) $^{40}$Ar$^+$ ions was regularly stored in CSR. The detector was routinely switched on to detect the neutral Ar products, originating from residual-gas electron capture by Ar$^+$, in order to deduce the stored-ion lifetime \cite{csrpaper}. Additionally, the detector was irradiated by 245-nm photons from the UV-LED for comparison of the signals (see below).\par 
Like the heavy particles, the UV photons do not irradiate the MCPs directly, but release electrons from the surface of the converter electrode which are then accelerated towards the MCPs. During UV irradiation, the aluminium converter thus acts as a photo-cathode of low ($\sim 10^{-4}$) quantum efficiency \cite{Dowell}. It was verified that, when the electron acceleration potentials were disabled while keeping the MCP gain voltage enabled, the count rate of the set-up dropped to zero. This shows that the MCP indeed detects secondary electrons only, and no primary particles (photons or ions) reach it in normal operation. Via the driving voltage of the UV-LED, the rate of photon detections could be varied over many orders of magnitude. In contrast to fast ions, each photon can emit at most only a single electron from the converter electrode, as the photon energy of 5.1(1)~eV is lower than the double work-function of the cathode material.\par
After pre-amplification, the pulses were discriminated and counted using the VME multiscaler, while sample waveforms were recorded by the oscilloscope. The MCP bias current was continuously measured by a floating nano-amperemeter. Figure~\ref{mcptemp} shows the derived electric resistance of the stacked MCP set as a function of the average temperature of the relevant CSR sector. No dedicated temperature sensor is attached to the particle detector itself. However, as the CSR cool-down process was slow, we assume that the MCPs were in thermal equilibrium with their surroundings.\par
Starting at the specified value of 56~M$\Omega$ at room temperature, the resistance of the chevron MCP-set rose by almost four orders of magnitude during the cool-down, reaching values of $\sim$\,300~G$\Omega$ at 6~K. After switching the detector on, a gradual increase of the MCP bias current by up to a factor of three was routinely observed within the first hour of operation, especially at very low temperatures. It is yet unknown whether this effect is due to operation-induced warming up or whether it reflects a purely electric change in the channel-plate properties. Figure~\ref{mcptemp} shows only the initially measured MCP resistance, directly after enabling of the high-voltage supplies, when the temperature of the plates can be assumed to have been equal to the temperature of the CSR vacuum chamber.\par
\begin{figure}[tb]
 \centering
 \includegraphics[width=\columnwidth]{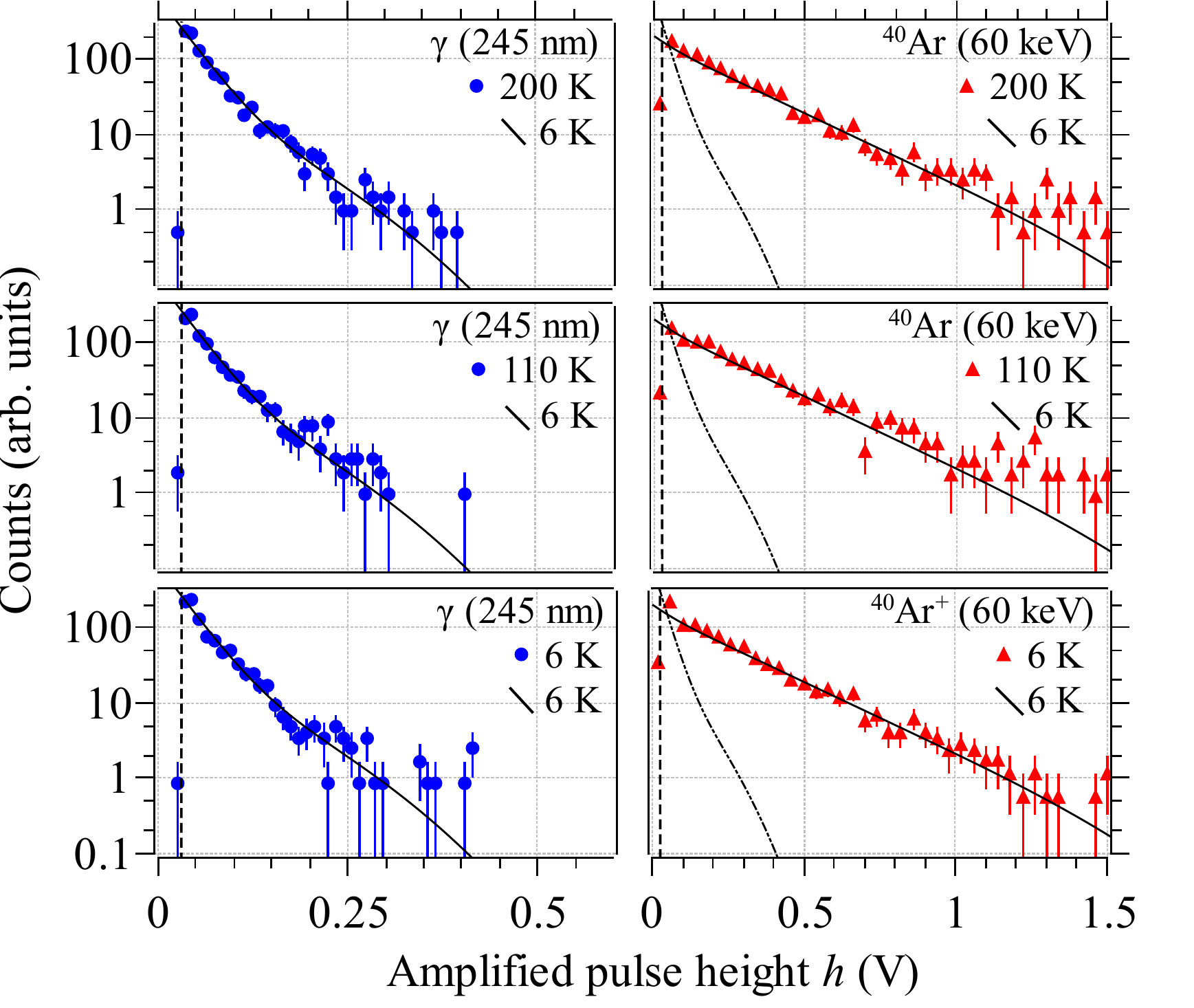}
 \caption{Pulse height distributions of the amplified detector pulses at three different temperatures of the CSR vacuum chamber (indicated by the labels 200~K, 110~K, and 6~K). The detector was irradiated alternatively by 245-nm UV photons (5.1(1)~eV, filled circles) and by 60-keV argon atoms ($^{40}$Ar and $^{40}$Ar$^+$, filled triangles). The pulse height spectra are normalised to equal total numbers of counts. The vertical dashed lines indicate the discrimination threshold of $\sim$\,35~mV. To ease comparison, a fit to the pulse height distribution at 6~K is shown in each frame of the respective particle (solid lines, cf.\ Sect.\ \ref{efficiencymodel}). To emphasise the different horizontal scales, the UV pulse height fit is also shown in the argon plots (dash-dotted lines).\label{coolpulses}}
\end{figure}
Figure~\ref{coolpulses} shows the pulse height distributions of the amplified detector signals obtained for UV photon and 60-keV argon irradiation at three different temperatures during the cool-down of CSR. The gain voltage across both MCPs was kept constant at 1.85~kV in all measurements. Pulses were recorded above a discrimination threshold of $\sim 0.035$~V. For comparison, the individual measurements were normalised to the number of counts. No significant changes in the distributions were found, neither for UV photons nor for 60-keV argon particles, between room temperature and the final operating point of CSR of 6~K.\par 
The intensity of the UV light source was set such that the average rate of detected photons was $\sim 600$~s$^{-1}$ in each measurement. For argon irradiation, the detected particle rate varied strongly with temperature, as---at given intensity of the stored ion beam---the rate of electron capture events scaled with the residual gas pressure in CSR, which improved drastically during the cool-down \cite{csrpaper}. At 200~K the detector recorded neutral Ar products at rates up to several $10000$~s$^{-1}$ while at 110~K, the production rate had decreased to a few $1000$~s$^{-1}$. At the final CSR temperature of 6~K, the rate of electron capture products was too low to be identified above a low background event rate of $\sim 10$~s$^{-1}$, in spite of the high stored ion current of $\sim 1$~$\mu$A. As the background events were not localised to the position of the axis of a neutral daughter beam, they are believed to be due to stray secondary particles produced along the beam pipe by the primary ions. In order to obtain a reliable pulse height distribution from impact of 60-keV argon particles at 6~K, the detector was moved towards the closed orbit in the storage ring until direct hits from parent Ar$^+$ ions could be detected at a rate of $\sim 500$~s$^{-1}$. It was assumed that the secondary electron ejection coefficient of 60-keV Ar$^+$ ions was sufficiently similar to that of neutral argon atoms of the same energy. Indeed, the measured pulse height distribution of Ar$^+$ ions corresponded to that of the neutral atoms at higher temperatures, as shown in Fig.~\ref{coolpulses}.            
\subsection{Localised Heating of the MCPs\label{heating}}
With the beam guiding vacuum chamber of CSR at 6~K, the electric heater built into the detector can be used to warm up the micro-channel plate set above the temperature of its surroundings. A functional test showed that, by operating the heating wire at a power of $\sim 80$~mW, the resistance of the MCP stack could be lowered by a factor of $\sim 10$. An approximate calibration, as indicated in Fig.~\ref{mcptemp}, translates this change in resistance to a warming of the channel-plate set from 6~K to $\sim 15$~K. No warming of the neighbouring CSR structures was observed in the process. Previous experiments have shown that even substantially greater heating powers can be applied without danger to the detector \cite{spruck}.\par 
The pulse height distribution obtained for UV irradiation did not change during the tests of the heating. This confirmed the earlier observations, as the UV-induced detector signals had also not been influenced by the cooling-down from room temperature (cf.\ Fig.~\ref{coolpulses}). It is however expected that the option of localised heating of the MCPs can improve the detector response to high-rate impact of massive particles (cf.\ Sect.\ \ref{performance}). In the experiments reported in the following, which focussed on the extreme low-temperature behaviour of the device, this possibility was not yet checked, and the MCP stack was deliberately left at the 6~K temperature of the surrounding CSR vacuum chamber. 
\subsection{General Performance at 6~K \label{performance}}
\begin{figure}[tb]
 \centering
 \includegraphics[width=\columnwidth]{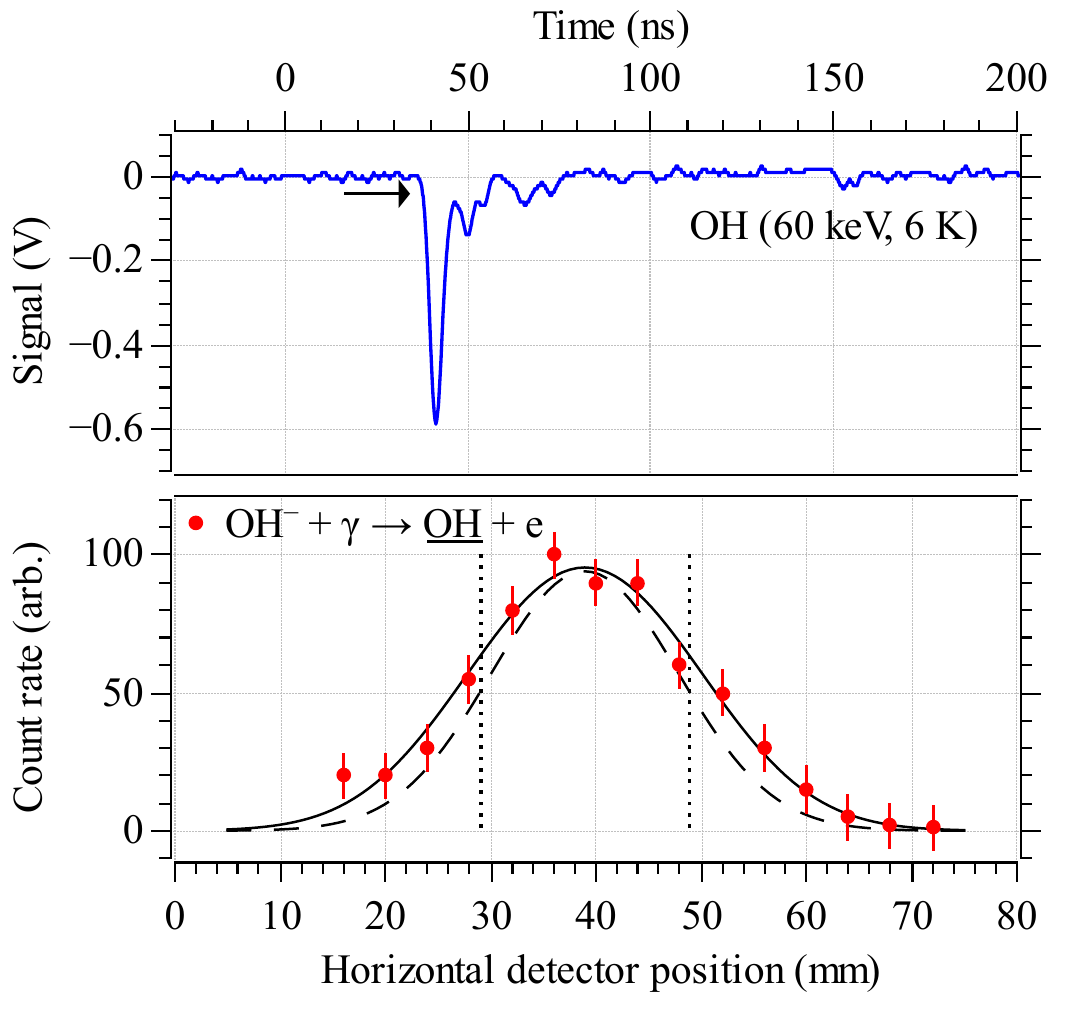}
 \caption{Sample amplified detector pulse obtained from a neutral product beam (top frame, solid line) and average discriminated count rate as a function of the horizontal detector position (bottom frame, filled circles) measured at a temperature of 6~K. The detected particles were neutral OH molecules of 60~keV total kinetic energy, originating from photo-detachment of stored OH$^-$ anions. The arrow indicates the discrimination threshold of $\sim 37$~mV. The solid line in the lower frame is a fit of an assumed Gaussian daughter beam envelope (dashed line, renormalised), convolved with the known horizontal detector aperture of 20~mm (indicated by the vertical dotted lines). \label{posscan}}
\end{figure}

With the CSR operating at its lowest temperature, the COMPACT set-up was employed to detect a variety of neutral and charged product beams. For single-particle experiments, the stored beam currents were well below that of the Ar$^+$ ions used for CSR commissioning. The above mentioned secondary-ion background was not observed in any other experiment, and even weak daughter beams could be easily identified by moving the detector horizontally across the CSR beam line and monitoring the average particle count rate as a function of travel distance, as shown in Fig.~\ref{posscan}. During the first experimental campaign, the cryogenic thread drive---as described in ref. \cite{spruck}---was used to move the detector by a total distance of more than $\sim 7$~m at lowest temperature, equivalent to 24 full strokes across the CSR vacuum chamber.\par
In the example of Fig.~\ref{posscan}, a product beam of 60-keV neutral OH molecules (3.53~keV/u) was detected, originating from electron detachment of stored OH$^-$ ions in a 633-nm cw laser beam in the experimental CSR section preceding the COMPACT detector. Using the known size of the horizontal detector aperture of 20~mm and the particle count rate measured as a function of detector position, the horizontal transverse daughter beam envelope can be obtained by deconvolution. The standard deviation of the assumed Gaussian product beam profile was derived to be 9.0\,(5)~mm at the detector position. As the neutral particles are not influenced by the ion optics, and as the momentum transfer to the molecule during the photo-detachment is negligible, the product beam maintains the emittance of its parent. Using the horizontal beta function of the storage ring \cite{csrpaper} one derives a 95\% horizontal transverse emittance of the ion beam of 24\,(3)~mm~mrad. One also derives that, in this example, the horizontal width of the daughter beam leads to a 27\,(3)\,\% geometric detection loss due to the narrow sensitive aperture. This is by design and not considered critical, as the detector is intended primarily to detect products originating from future electron-cooled ion beams \cite{spruck}. Such beams are characterised by a much lower transverse emittance, and in their case the narrow detector aperture will help identifying products based on charge-to-mass selection. Note that the height of the sensitive aperture is much wider ($50$~mm \cite{spruck}) so that no significant vertical cut on the product beam is believed to occur in the experiments described here. Variants of the COMPACT detector with larger horizontal acceptance for use in future measurements on uncooled ion beams are presently being developed.\par 
\begin{figure*}[tb]
 \centering
 \includegraphics[width=\textwidth]{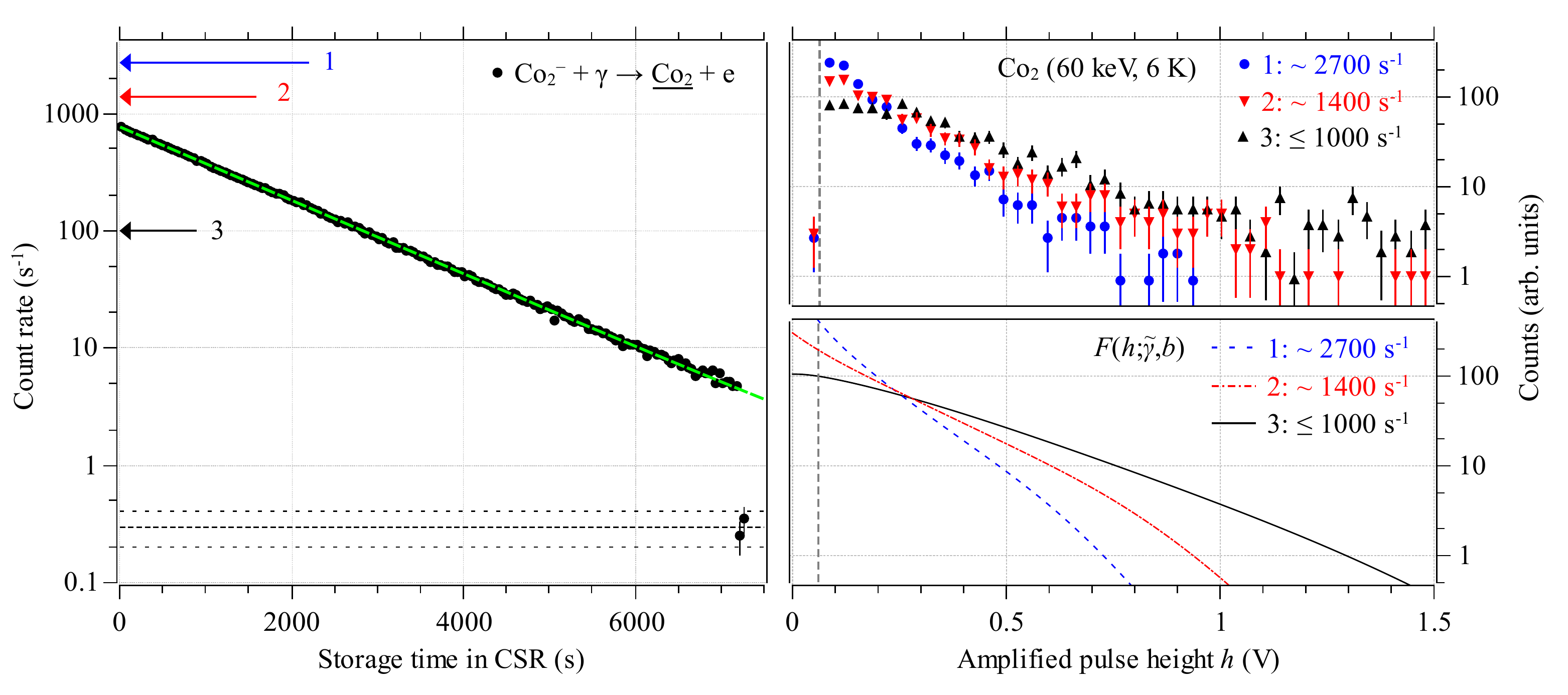}
 \caption{Detected product rate from photo-detachment of Co$_2^-$ anions as a function of storage time in CSR at 60~keV total energy (0.51~keV/u) and 6~K operating temperature (left frame), and pulse height distributions obtained from the 60-keV Co$_2$ molecules at different average hit rates (right frames). The distributions 1--3 (see annotation) were measured at the indicated average count rates. The right bottom frame shows fits of the measured pulse height distributions using the model from Sect.~\ref{efficiencymodel}, including damping of the MCP gain at high rates in the cases 1 and 2 (see text). Below $\sim 1000$ counts per second, no dependence of the pulse heights on the detection rate was observed. The left frame shows the discriminated count rate (dots) of neutral Co$_2$ products from photo-detachment of a suitable 60-keV Co$_2^-$ beam, stored in CSR for two hours. After 7200~s of storage, the ion beam was deliberately kicked out of the closed orbit. The long-dashed line is an exponential decay fit, yielding a $1/e$ lifetime of the stored ions of 1383(5)~s. The horizontal short-dashed lines indicate the intrinsic dark count rate of the detector, which was found to be as low as 0.3(1)~s$^{-1}$ in all experiments described here.\label{rates}}
\end{figure*}
The detector pulse shapes and height distributions found for atomic or molecular products from experiments at 6~K were quite similar to those observed for argon hits during cool-down (cf.\ Fig.~\ref{coolpulses}). As shown in Fig.~\ref{posscan}, the amplified anode signals are 10--30~ns short pulses of typically a few 100~mV amplitude.\par
Characteristic features found in all experiments conducted during CSR commissioning are pulse height distributions which are not peaked (cf.\ Figs.\ \ref{coolpulses} and \ref{rates}). The explanation for this signature lies in the fact that the chevron MCPs detect the 1.2-keV secondary electrons ejected by the primary ions \cite{spruck}. All of these secondary electrons have to be assumed to impinge close to \emph{different} MCP channels, within a time shorter than the observed pulse width of $\sim 10$~ns, as has been verified by numerical simulation of the electron trajectories in the detector. Hence, the total pulse height from heavy-ion impact is, in fact, the result of pile-up of several independent pulses generated by 1.2-keV electron impact on the MCPs. The resulting sum pulse height distribution tends to be monotonously decreasing with higher amplitudes, as will be discussed in the upcoming Sect.~\ref{results}, following the original analysis by Spruck et al.\ \cite{spruck}. Due to this characteristic pulse height distribution, the detector count rate was sensitive to the signal discrimination threshold in the present work, and a low electric base-line noise was imperative in order to obtain good overall detection efficiency.\par 
\subsection{High-Average Rate Response at 6~K\label{Codcrate}}
At 6~K, a dependence of the pulse heights on the average detector count rate was observed for heavy-particle impact above a certain critical hit rate. This is illustrated in Fig.~\ref{rates}: In the experiment, a beam of Co$_2^-$ ions was stored in CSR at 60-keV total energy (0.51 keV/u) and passed through a grazing-angle, 633-nm-laser target in the experimental section. Photo-detachment yielded neutral Co$_2$ molecules that reached the detector at the kinetic energy of the parent beam. By variation of the intensities of the ion or laser beams, the average rate of product particles could be adjusted.\par 
It was observed that above $\sim 1000$ discriminated Co$_2$ hits per second in average, the pulse amplitudes started to decrease noticeably (cases 1 and 2 in Fig.~\ref{rates}). At a detection rate of $\sim 2700$~s$^{-1}$ (case 1) the mean pulse amplitude was $\sim 25$\,\% lower than below 1000~s$^{-1}$ (case 3). Given the simple counting logics based on a fixed signal discrimination threshold (dashed vertical line in Fig.~\ref{rates}), this caused the detected particle rate to vary non-linearly with the stored intensity of the Co$_2^-$ beam above a discriminated count rate of 1000~s$^{-1}$. In contrast, UV photons from the test LED---which produce much smaller MCP signals---could be detected at 6~K at significantly higher rates (more than $\sim 3000$~s$^{-1}$) without deterioration of their pulse amplitudes. We attribute this effect to gain saturation due to the onset of charge depletion of the MCPs at simultaneously low temperature and elevated heavy-particle hit rate. This hypothesis is also supported by the earlier observation that, at higher temperature, 60-keV Ar could be detected at much higher average count rates with no evidence of signal degradation (c.f.\ Fig.~\ref{coolpulses}). It is expected that local warming of the MCP set (cf.\ Sect.\ \ref{heating}) can mitigate these saturation effects, however this has not been studied yet.\par
In the Co$_2^-$ experiment the pulse height distribution---and thus the discriminator efficiency---was rate-independent also at 6~K, as long as the average count rate was kept below 1000 hits per second. Under these conditions, the detector could be used to reliably measure the evolution of the photo-detach\-ment rate over very long storage times of the Co$_2^-$ ions (left-hand frame of Fig.~\ref{rates}). A fit to the Co$_2$ count rate as a function of storage time yields a $1/e$ lifetime of the anions in the experimental set-up of 1383\,(5)~s. After 7200~s of storage, the remaining ions were deliberately kicked out of their closed orbits within a single turn, so that the dark count rate of the detector could be measured. As visible in the left-hand frame of Fig.~\ref{rates}, even two hours after ion injection, the measured Co$_2^-$ photo-detachment rate was still more than an order of magnitude greater than the detector background level. In all measurements, the latter was found to be 0.3\,(1)\,s$^{-1}$, with no notable dependence on temperature. Most of the dark pulses are believed to be due to $\beta$-decay of radio-nuclides in the MCP substrate \cite{siegmund} and are very similar to actual counting pulses in shape and amplitude. The fact that this background rate is found to be very low is important in that rejection of the dark events based on pulse shape analysis does not at present seem feasible.\par   
\subsection{Short Particle Bursts at 6~K}
In the experiments described so far the product particles reached the detector in quasi-steady streams, with average fluxes that varied slowly compared to all other time constants of the set-up. Due to the long storage times of the CSR, this situation is common for parent ions that are in a stable state, or when their internal state population varies slowly, such that the rate of interaction between the stored beam and the target is nearly constant.\par   
In other cases the reaction products show a burst-like time-structure. E.g., interaction of the stored ions with a pulsed laser target or the ion production process itself can lead to population of metastable levels. By timing the detection of the resulting end products with respect to the time of interaction, the lifetimes of the metastables can be measured down to the scale of the revolution period in the storage ring. In such experiments, the detector hit rate may vary drastically within a few milliseconds, with the burst rate directly following the formation of the metastables significantly exceeding the average count rate in the experiment.\par
As an example, the auto-fragmentation of Co$_2^-$ molecular ions was studied. The Co$_2^-$ beam was produced in a metal-ion sputter source, accelerated to a total kinetic energy of 60~keV, and stored in CSR for 90~s. After that storage time, any remaining ions were dumped before the next injection took place. Also here, the storage ring operated at 6~K. The sputter ion source naturally produces part of the anions in auto-dissociating metastable states. The COMPACT detector was positioned such as to collect the 30-keV Co$^-$ fragments that, in absence of residual gas collisions or other target interactions, could only be produced from the metastable ion population. The results of the experiment will be published separately.\par
\begin{figure}[tb]
 \centering
 \includegraphics[width=\columnwidth]{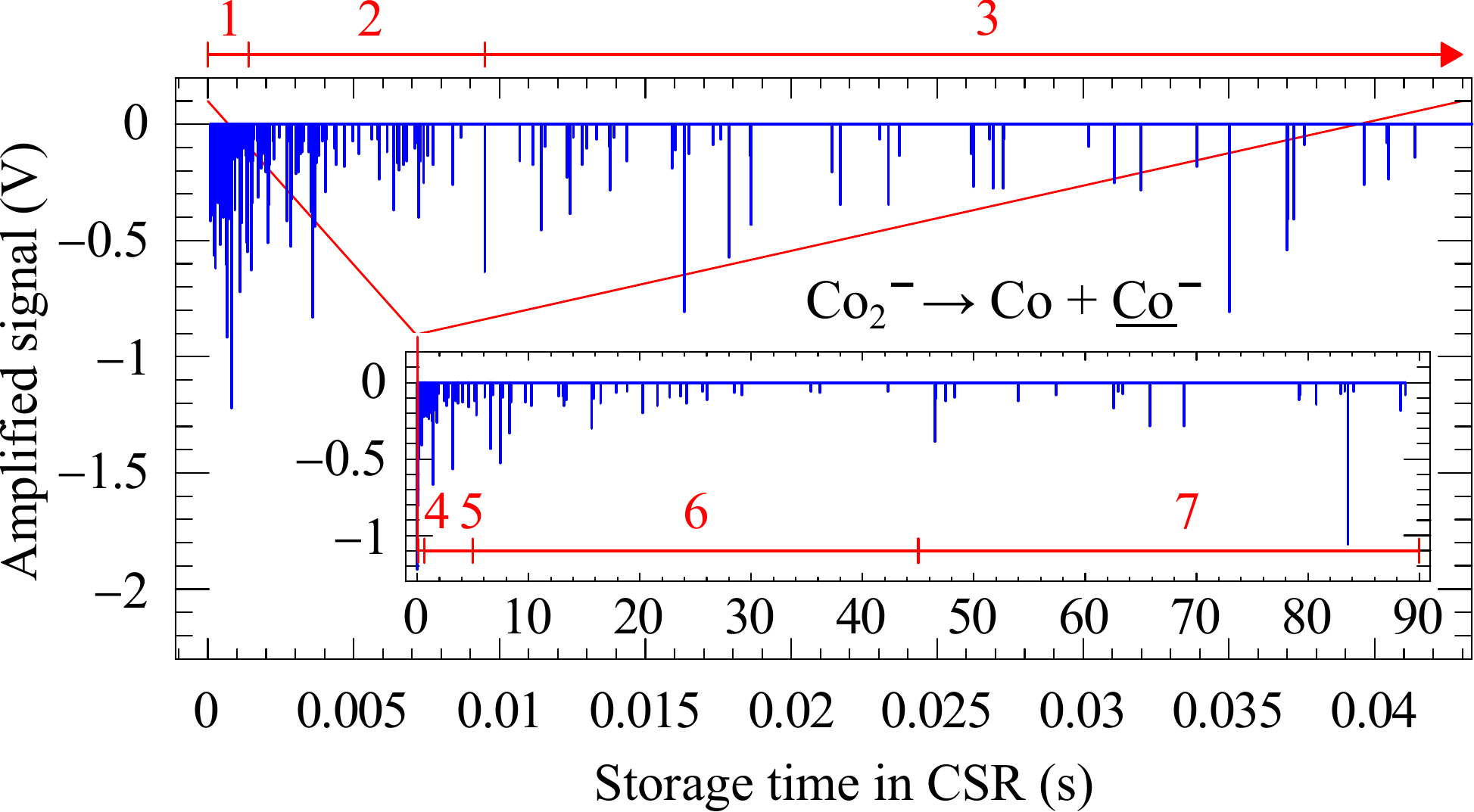}
 \caption{Detector pulses generated by Co$^-$ anions from auto-dissociation of Co$_2^-$ molecules stored in CSR. The MCP set was operating at the 6~K temperature of the surrounding beam-line. The data-set corresponds to a single ion-injection into the storage ring. Due to the short lifetime of the excited Co$_2^-$ molecular ions, the rate of Co$^-$ hits steeply decreased after the start of the measurement. The evolution of the detector signals was observed as a function of time (and thus average count rate) by evaluating the mean pulse height distributions in the indicated time intervals.\label{Copulses}}
\end{figure}
\begin{figure}[tb]
 \centering
 \includegraphics[width=\columnwidth]{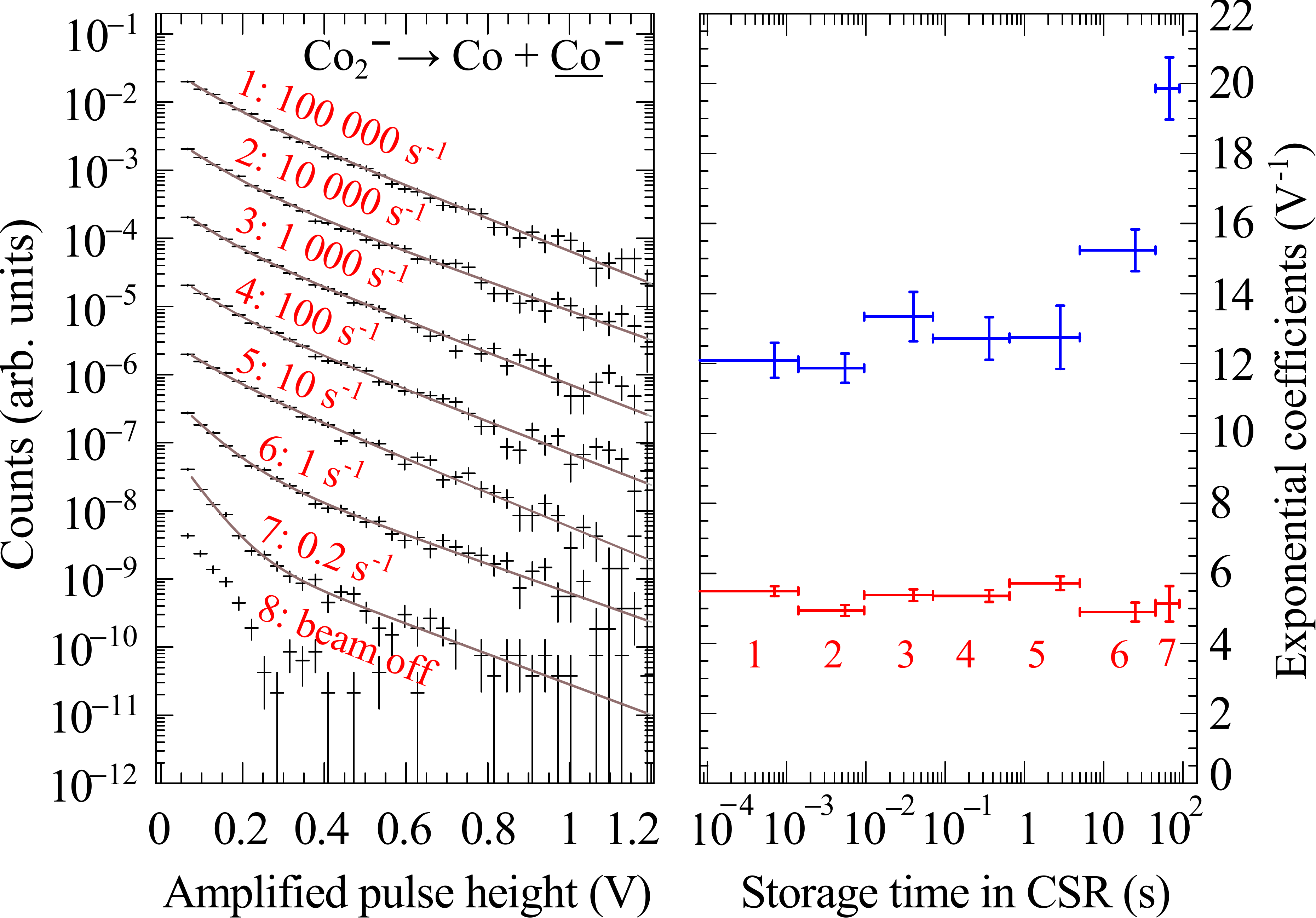}
 \caption{Left: Pulse height distributions obtained for 30-keV Co$^-$ anions reaching the detector following auto-fragmentation of stored Co$_2^-$ molecular ions. The labels correspond to the time intervals as indicated in Fig.~\ref{Copulses}. The distributions are artificially shifted by multiplication by factors of ten for better readability. The labels indicate the approximate average count rate in the respective interval. Right: Decay coefficients obtained from double exponential decay fits of the pulse height distributions for the different time bins. The change in the later pulse height distributions 6 and 7 is due to the rising contribution of detector dark counts (distribution 8) to the measured signal.\label{Cophds}}
\end{figure}
As shown in Fig.~\ref{Copulses}, the instantaneous detection rate of Co$^-$ was very high directly after ion injection, but then steeply decreased on a time-scale of milliseconds. In the experiment, the advanced FADC-based data acquisition system (cf.\ Sect.~\ref{experimental}) was used. In contrast to the steady-current experiments, as described in Sect.~\ref{Codcrate}, it was thus possible to observe changes in the pulse-height distribution on very short time scales, limited only by the counting statistics.\par 
Starting with the ion injection into CSR, integration time windows were defined as shown in Fig.~\ref{Copulses}. For each integration window, the height distribution of the detector pulses and the average count rate were calculated. The length of the time windows increased as a function of storage time, so that each window was characterised by the approximate average count rates given in Fig.~\ref{Cophds}. To easily quantify the shape of the measured pulse height distributions, each was approximated by a simple double exponential decay fit function.\par
As visible in Fig.~\ref{Cophds}, at 6~K detector temperature the pulse height distribution was constant across storage time, even if the peak rate during the first millisecond after ion injection was as high as $10^5$~s$^{-1}$, i.e., two orders of magnitude higher than the maximum useful count rate observed in the Co$_2^-$ photo-detachment experiments described in Sect.~\ref{Codcrate}. The apparent change in shape at very late storage times (distributions 6 and 7 in Fig.~\ref{Cophds}) is due only to the increasing contribution of detector dark counts (distribution 8) to the measured signals.\par 
The different saturation threshold compared to the (steady-current) photo-detachment experiment from Sect.~\ref{Codcrate} is most likely due to the fact that the short auto-fragmentation bursts were followed by extended periods of near-zero count rate, during which the MCP channels could recharge via the low bias current at 6~K, before the next ion injection would take place. For peak hit rates even larger than $10^5$~s$^{-1}$, saturation effects, similar to those depicted in Fig.~\ref{rates}, were indeed  observed also in the burst-type auto-frag\-ment\-ation experiment. In those cases, the non-linearity in the measured count rate with respect to the true fragment production rate could not be fully eliminated by data processing, in spite of the FADC-based data acquisition system allowing for advanced pulse discrimination techniques. Also in the burst experiments, on-line measurement of the pulse height amplitudes along with the detection rate thus turned out to be a crucial tool for evaluating the reliability of the experimental data.
\section{Analysis\label{results}}
In the following we seek to quantify the experience from the first operation of the COMPACT detector at 6~K temperature. The many engineering topics related to the cryogenic environment have been discussed in detail by Spruck et al.\ in a previous article \cite{spruck}. Here we focus on the performance of the instrument during the first atomic and molecular physics experiments using the storage ring CSR at lowest temperature. For a particle counting detector, two basic properties come to mind: They are the single particle detection efficiency and useful dynamic range of the count rate.\par
\subsection{Dynamic Range}
The dynamic range is defined by the intrinsic detector background on the one hand, and by the maximum particle count rate that can be reliably measured on the other hand. For the chevron MCP set operating at the lowest CSR temperature of 6~K, the experiment on Co$_2^-$ photo-detachment from Sect.~\ref{Codcrate} can be considered as a benchmark: from the dark event rate of 0.3\,(1)\,s$^{-1}$ to the maximum product count rate of $\sim 1000$~s$^{-1}$ that could be reliably discriminated, the dynamic range of $\sim 3\times10^3$ in continuous-rate measurements allows to follow product formation from atomic, molecular or cluster processes in the CSR for up to eight $1/e$-lifetimes of the reaction at hand. It is expected that local heating of the MCPs can extend the dynamic range even further, but this has not been studied yet. For burst mode operation, the Co$_2^-$ auto-fragmentation measurement---also carried out at 6~K detector temperature---shows a significantly greater dynamic range, reaching up to $\sim 3\times10^5$ in the example at hand. However, that value likely depends on the time the MCP set is allowed to recharge in-between the bursts of product particles, an effect that has not been studied systematically yet.\par
This favourable low-temperature behaviour of the detector is believed to be a consequence of its `Daly'-type design. The detector combines a large sensitive aperture---defined by the size of the `Daly' converter electrode---with a relatively small MCP set that collects the secondary electrons ejected from that electrode. The background event rate of MCPs naturally scales with the volume of the substrate \cite{siegmund}. An EDR-MCP of the same active area as the sensitive aperture of the COMPACT detector ($20 \times 50$~mm) can be expected to have a dark count rate that is an order of magnitude higher than the one measured in the experiments reported here. Additionally, larger (and thicker) MCPs have been measured to reach much higher electric resistance near liquid-helium temperature than the $\sim 300$~G$\Omega$ found here \cite{schecker}. Even EDR variants of large channel-plates have been found with unfavourable electric behaviour at lowest temperature, which likely leads to earlier depletion at high detection rates and thus worse high-rate acceptance \cite{roth,kuehnel}. Small MCPs are also produced on a large scale routinely, which---besides the obvious advantage of lower prices---might lead to more stable production processes and more predictable properties. 
\subsection{Detection Losses\label{losses}}    
Knowledge of the particle detection efficiency can be important in experiments seeking to measure absolute cross sections of ion reactions in the storage ring. In some cases the detector can be calibrated against a known process, or its efficiency can be inferred by controlled variation of the experimental conditions. However, if other parameters of the experiment are unknown, independent knowledge of the product detection efficiency can be the only way to interpret the measurements in terms of absolute numbers. Cryogenic operation of MCPs is considered out-of-specification by their manufacturers, and the detector behaviour near liquid-helium temperature is not guaranteed. Although there has been some research on the topic, open questions remain \cite{schecker,roth,kuehnel,rosen}. In this situation, a simple way to monitor the absolute detection efficiency during the experiment is important to ensure the reliability of the data taken.\par 
In the experiments reported here, the particle detection efficiency of the COMPACT detector is limited by three effects. The first is the loss of product particles due to limited geometric acceptance of the detector in the horizontal plane (cf.\ Fig.~\ref{posscan}). This is not discussed further here. As noted above, the narrow width of the sensitive aperture is by design. In fact significant efforts have been undertaken to realise the vertically elongated detection window. It is still wide enough to intercept daughters of electron cooled beams with 100\% efficiency \cite{spruck}. In all other cases the geometric loss ratio can be determined by a horizontal scan of the daughter beam envelope as described in Sect.~\ref{operation}.\par 
A second limitation of the detection efficiency arises from the discrimination threshold applied to the anode signals. Pulses of amplitudes below threshold are not recorded in the counting electronics used in the experiments presented here. This is an issue that must be addressed technically. At high gain of the pre-amplifier and simultaneously low baseline noise level of the anode high-voltage line, the discrimination threshold can be very low relative to the mean pulse height. The remaining, small cut-off ratio can be easily estimated if the overall shape of the pulse height distribution is known or can be extrapolated. FADC-based data acquisition systems, like the one presented in Sect.~\ref{experimental}, may not involve a fixed discrimination threshold at all, as they allow identification of particle hits using numerical pulse-shape analysis of the time-resolved anode signal. Independent external triggers (from, e.g., a pulsed laser or the storage-ring timing system) are then typically used to start and stop the FADC. Whether the FADC data can be processed in real-time, or needs to be recorded for subsequent analysis, depends on the processing speed of the computer, the particle rate, and the complexity of the chosen pulse detection algorithm.\par
The third and most fundamental source of detection efficiency loss lies in the stochastic nature of the electron ejection process from the converter cathode and of the detection of these secondary electrons by the MCPs. Even if the average number of electrons released per impinging ion can be quite high in some experiments, there is in fact a non-zero probability that an ion either releases no electron at all, or that none of the ejected electrons is detected by the MCPs. In these cases no anode pulse can be observed, no matter how technically advanced the readout electronics is. In the following, we denote by $P_0$ the likelihood for no MCP multiplication event to occur, although the converter electrode \emph{did} receive an impact from a heavy particle.\par
\subsection{Modelling the Detection Efficiency\label{efficiencymodel}}
In the case of the COMPACT detector, a value of $P_0$ can be derived by comparison of the anode pulses obtained for the heavy particle under study with those generated by UV photons from the LED source installed in CSR. As discussed in Sect.~\ref{operation}, anode pulses for heavy-particle impact can be assumed to result from pile-up of the MCP signals generated by the secondary electrons released from the `Daly' converter cathode. In contrast, the 245-nm photons can be assumed to release at maximum a \emph{single} electron from the cathode material due to their low energy (5.1(1)~eV). Comparison of the pulse height spectra obtained in both cases thus allows to estimate the average number of secondary electrons contributing to the heavy-ion signals.\par    
In the case of the experiment on stored Ar$^+$ shown in Fig.~\ref{coolpulses}, comparison of the mean pulse heights obtained for 60-keV argon and UV irradiation suggests that, in average, 4--5 MCP events from secondary electrons pile up to form the Ar-induced signals. At an assumed MCP detection efficiency of 60\% for the 1.2-keV electrons, this means that for each impinging heavy Ar atom an average number of $\tilde{\gamma} \approx 7.5$ secondary electrons reach the MCP surface. Based on Poisson-statistics, one would expect the chance for \emph{no} charge multiplication to occur in the MCP-stack after a heavy-ion impact to be as low as $\sim 1$\%.\par      
The Poissonian model is however only true if the point of impact of the ions on the detector---and thus the average number $\tilde{\gamma}$ of converter electrons attracted towards the MCP surface---is fixed \cite{spruck}. This is expected to be a good approximation in future experiments on electron-cooled atomic ions in CSR, as their product beams are characterised by very low emittances. In absence of beam cooling---as in all experiments reported here---or for strongly exothermic molecular breakup reactions, the product particles irradiate a large fraction of the sensitive aperture of the detector. In that case a possible variation of $\tilde{\gamma}$ across the sensitive detector aperture must be accounted for.\par   
A model of the secondary-electron statistics valid for non-uniform $\tilde{\gamma}$ has been developed within the framework of discrete-dynode electron multipliers, where a similar situation occurs \cite{prescott}. There, the number $n$ of electrons emitted from one dynode towards a second one is described by a Pólya distribution  
\begin{equation}
 W_n(\tilde{\gamma},b) = \frac{\tilde{\gamma}^n}{n!}(1+b\,\tilde{\gamma})^{-n-1/b}\ \prod_{j=0}^{n-1} (1 + j\,b)\ . 
 \label{polya}
\end{equation}
$\tilde{\gamma}$ is now the mean number of secondary electrons reaching the second dynode for each impact on the first one. $b$ is the relative variance of $\tilde{\gamma}$. For $b=0$, $W_n$ is equal to the Poisson distribution, in the special case of $b=1$ it assumes the shape of the exponentially decreasing geometric distribution. In our application we identify the `Daly' converter cathode with the emitting dynode, while the role of the collecting dynode is assumed by the positively biased MCP input surface.\par   
\begin{figure}[tb]
 \centering
 \includegraphics[width=\columnwidth]{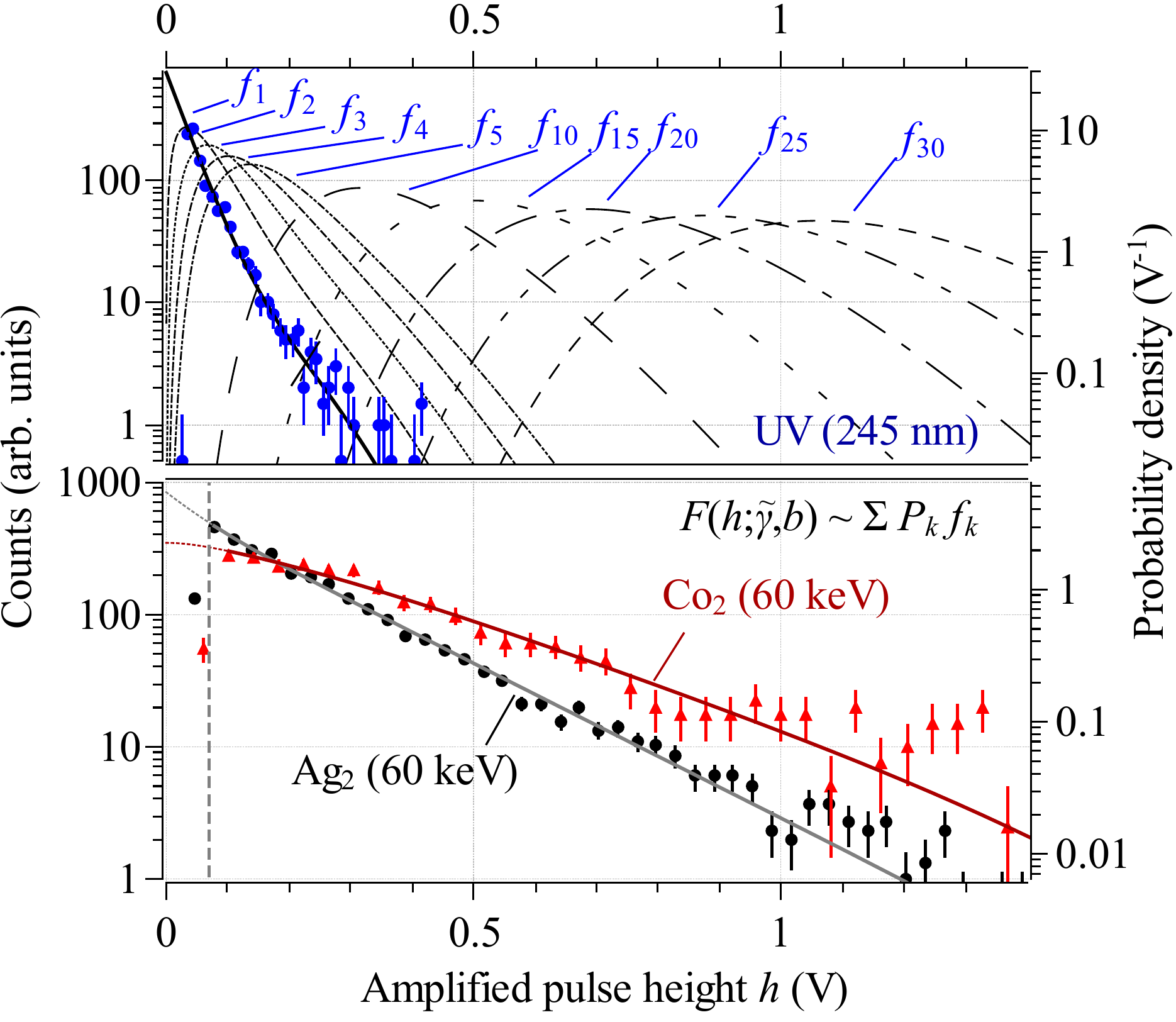}
 \caption{Illustration of the fit procedure for determination of the detector efficiency. Top frame: Pulse height distribution ($f_1$) measured for 245-nm photon irradiation of the detector at 6 K temperature (filled circles) and fit (solid line). The numerically computed convolutions $f_k$ (for $k>1$) are indicated by the fine dashed and dash-dotted lines as indicated by the labels. Bottom frame: Via Eq.~(\ref{fitfun}) the $f_k$ sum up to the pulse height distributions for heavy-particle impact ($F$, see text). Best-fit models of $F$ are shown for 60-keV Ag$_2$ and Co$_2$ products (solid lines, as labelled), reaching the detector at average rates of $\sim 100$~s$^{-1}$ and 6 K operating temperature.\label{fits}}
\end{figure}
Each of the $n$ electrons from Eq.~(\ref{polya}) has a chance $\epsilon$ to generate a charge multiplication avalanche in the MCPs. Hence, the total number $k$ of MCP cascades generated by a single primary ion is distributed according to
\begin{equation}
 P_k(\tilde{\gamma},b)  = \sum_{n=k}^\infty \binom{n}{k}\epsilon^k(1-\epsilon)^{n-k} \,W_n(\tilde{\gamma},b)\ ,
      \label{pulseheightpolya}
\end{equation}
which is the discrete convolution of $W_n$ with a binomial distribution. In the following, we assume $\epsilon = 0.6$ based on the geometric open-area ratio of the MCPs. The principal results of the analysis are largely independent on the choice of $\epsilon$ as discussed below.\par
Let $f_k(h)$ be the distribution of pulse heights $h$ produced by the MCP-stack upon simultaneous multiplication of (precisely) $k$ converter electrons. If $f_k$ is known for all $k$, the sum pulse-height spectrum $F$ for heavy-particle detection can be modelled as   
\begin{equation}
 F(h;\tilde{\gamma},b) = C\,\sum_{k=1}^\infty P_k(\tilde{\gamma},b)\,f_k(h) \ ,
 \label{fitfun}
\end{equation}
with $C$ being a normalisation factor. In the case of the COMPACT set-up, the $k$-electron distributions $f_k$ can be inferred from the pulse height distribution $f_1$, measured for irradiation of the detector by the UV photon source. As the photons never emit more than one converter electron, their MCP pulse height distribution is equal to $f_1$. The pulse height spectra for $k$-fold pile-up signals are then given by the recursive convolution formula $f_k = f_1 \ast f_{k-1}$ (for $k > 1$).\par
With all $f_k$ known, $\tilde{\gamma}$ and $b$ can be obtained from a fit of Eq.~(\ref{fitfun}) to the data of each given CSR experiment. Via Eq.\ (\ref{pulseheightpolya}), these parameters yield an independent experimental value for the likelihood $P_0$ and, hence, for the maximum possible detector efficiency due to secondary electron statistics, $1 - P_0$. Analytically, the normalisation factor $C$ from Eq.~(\ref{fitfun}) is equal to $(1-P_0)^{-1}$. However, due to the experimental discriminator cut-off, the measured pulse height distribution for heavy-particle impact cannot be reliably renormalised to 1 as long as the best-fit distribution $F$ is not known. For simplicity, $C$ is therefore treated as an independent free parameter.\par 
\begin{table*}[tb]
  \centering
  \begin{minipage}{0.9\textwidth}
    \centering
    \caption{Detection efficiencies derived from pulse height distributions measured in various CSR experiments using the probabilistic model from Eq.~(\ref{fitfun}). The second and third columns indicate the detected product and its specific kinetic energy. The fourth column shows the temperature of the detector during the experiment. In the measurements at lowest temperature (6~K), the average count rate was not higher than $\sim 300$~s$^{-1}$, so that distortion of the pulses due to charge depletion of the MCP-set should not have occurred. The fifth and sixth column show the best-fit values of the Pólya parameters $\tilde{\gamma}$ and $b$ (cf.\ Eq.~(\ref{polya})). The last three columns contain, respectively, the fraction of the fitted distribution observed above discriminator level, the theoretical maximum detector efficiency $1-P_0$ due to secondary-electron statistics, and the total detection efficiency for products reaching the detector. Note that particle loss due to the geometrical detector acceptance is not accounted for here.\label{fitresults}}
    \vspace*{1em}
    \begin{tabularx}{\textwidth}{ccccccccc}
      \hline
      \hline
      Reaction                                                        & Detected & Energy     &Temp.   & $\tilde{\gamma}$ & b       & $\ge$~disc. & $1-P_0$   & Total   \\
                                                                      & particle & (keV/u)    &$T$ (K) &               &            & (\%)        & (\%)      & (\%)    \\
      \hline
 $\mathrm{Ar}^+ + X \rightarrow \mathrm{Ar} + X^+$                    & Ar      & 1.50     & 200     &   7.5\,(6)    & 0.9\,(2)   & 84\,(1)     & 83\,(3)   & 70\,(3) \\
 $\mathrm{Ar}^+ + X \rightarrow \mathrm{Ar} + X^+$                    & Ar      & 1.50     & 110     &   8.7\,(8)    & 0.8\,(2)   & 85\,(1)     & 87\,(4)   & 74\,(3) \\
                    ---                                               & Ar$^+$  & 1.50     & 6       &   8.5\,(6)    & 0.8\,(2)   & 85\,(1)     & 87\,(3)   & 74\,(3) \\
      \hline 
 $\mathrm{N}_2^+ + X \rightarrow \mathrm{N} + \mathrm{N}^+ + X$       &  N$^+$  & 2.14     & 83      &   6.0\,(5)    & 0.5\,(2)   & 82\,(2)     & 88\,(4)   & 72\,(4) \\
 $\mathrm{N}_2^+ +X\rightarrow\mathrm{N}+\mathrm{N}^{(+)}+X+(\mathrm{e})$&N     & 2.14     & 83      &   10.6\,(6)   & 0.5\,(1)   & 90\,(1)     & 93\,(2)   & 84\,(2) \\
      \hline
 $\mathrm{O}^- + \gamma \rightarrow \mathrm{O} + \mathrm{e}$          & O       & 3.75     & 6       &   9.3\,(8)    & 0.8\,(2)   & 85\,(2)     & 87\,(4)   & 74\,(4) \\
 $\mathrm{OH}^- + \gamma \rightarrow \mathrm{OH} + \mathrm{e}$        & OH      & 3.53     & 6       &   8.8\,(6)    & 0.8\,(2)   & 80\,(2)     & 87\,(3)   & 69\,(4) \\
 $\mathrm{Co}_2^- + \gamma \rightarrow \mathrm{Co}_2 + \mathrm{e}$    & Co$_2$  & 0.51     & 6       &   12.1\,(7)   & 0.5\,(1)   & 83\,(2)     & 96\,(2)   & 80\,(3) \\
 $\mathrm{Ag}_2^- + \gamma \rightarrow \mathrm{Ag}_2 + \mathrm{e}$    & Ag$_2$  & 0.28     & 6       &   5.6\,(4)    & 1.2\,(2)   & 77\,(3)     & 73\,(3)   & 56\,(4) \\
      \hline
      \hline    
    \end{tabularx}
  \end{minipage}
\end{table*}
Figure~\ref{fits} illustrates the procedure: By UV-irradiation of the detector, we measure the single-electron spectrum $f_1$. The pile-up distributions $f_k$ are obtained by numerical convolution. The statistical sum spectrum $F$ from Eq.~(\ref{fitfun}) is then fitted to the pulse height distribution measured for heavy-ion detection, as Fig.~\ref{fits} shows on two examples.\par 
It should be noted that the method to extract $b$ and $\tilde{\gamma}$ from the detector pulse height distribution is not new. In fact it is the standard way to measure secondary electron yields of ions impinging onto solids \cite{dietz,lakits,collins,schackert,moshammer,itoh}. Normally a secondary electron detector with good energy resolution is used, so that the components $f_k$ show up as clearly resolved peaks in the measured pulse height spectrum. For the COMPACT MCPs, the pulse height spectrum $f_1$ for a single secondary electron (i.e.\ for UV irradiation of the detector) is found to be a monotonously decreasing exponential distribution. Unsurprisingly, the resolution with regard to electron multiplicity is therefore very bad. The aim of this analysis is not to derive the secondary electron yield $\tilde{\gamma}$ but to estimate the amount of undetected ions.\par
The assumed MCP electron detection efficiency $\epsilon$ was kept fixed at 0.6. Due to the effect of $\epsilon$ on the mean of the binomial distribution in Eq.~(\ref{pulseheightpolya}), its choice correlates inversely to the fit value of $\tilde{\gamma}$. The results for the detection efficiencies are however largely independent of that choice.\par 
In the fits of the pulse height distributions 1 and 2 from Fig.~\ref{rates}, showing the effect of detector saturation, $f_1$ was scaled by a factor $d<1$ along the $h$-axis, before the pile-up distributions $f_k$ were computed (cf.\ Eq.~\ref{fitfun}). This simulates the reduced gain of the MCP-detector with respect to unsaturated behaviour. For values of $d \approx 0.4$ (distribution 1 in Fig.~\ref{rates}) and $d \approx 0.6$ (distribution 2), the subsequent fit procedure yields values for $\tilde{\gamma}$ and $b$ that are compatible with the unsaturated case (distribution 3) within their statistical uncertainties.\par   
Table~\ref{fitresults} summarises the data from a few selected experiments. The first three rows show the analysis of the pulse height distributions found in the Ar$^+$ storage experiments during cool-down of CSR (cf.\ Sect.\ \ref{operation}), as shown in Fig.~\ref{coolpulses}. As noted earlier, no change in the pulse heights as a function of temperature was observed for either photon or argon irradiation of the detector. Also the fit results obtained using the model from Eq.~(\ref{fitfun}) are consistent among all three operating temperatures. The number of secondary electrons emitted by the converter cathode is derived as $\tilde{\gamma} \sim 8$, in good agreement with the above estimate obtained by comparison of the mean pulse amplitudes of UV-photons and heavy ions. The Pólya parameter $b$ fits at a large value of 0.8--0.9, which enhances the likelihood $P_0$ for emission of no secondary electron by a factor of $\sim 10$ compared to the Poisson-statistical case ($b=0$). The large value of $b$ may indicate that a large fraction of the sensitive detector aperture was irradiated by the Ar particles---as could be expected from an intense uncooled ion beam. The expected maximum possible detection efficiency $1-P_0$ is hence reduced to $86\,(3)\,\%$. The signal acquisition threshold causes another $\sim 15\,\%$ loss in efficiency, determined by the fraction of the best-fit model distribution (Eq.~(\ref{fitfun})) below discrimination level (cf.\ Fig.~\ref{coolpulses}). The resulting total detection efficiency for Ar atoms entering the detector is thus determined to be an average $73\,(3)\,\%$. Geometric loss of particles due to the narrow detector aperture---likely to have occurred in all experiments reported here---is not taken into account as it has been measured in the case of OH$^-$ only (cf.\ Fig.~\ref{posscan}).\par 
At intermediate temperature of CSR, dissociative residual-gas collisions of N$_2^+$ were observed. Significantly different pulse height distributions were found for the charged and neutral product beams (central two rows of Tab.~\ref{fitresults}), as reflected by the very different best-fit values of the secondary yield $\tilde{\gamma}$. Partly, the difference may be due to deflection of the charged fragments out of the storage ring plane, so that they hit a different area of the converter cathode. In addition, the neutral product beam is believed to originate not only from dissociative collisions $\mathrm{N}_2^+ + X \rightarrow \mathrm{N}^+ +\mathrm{N} + X$, but also from the competing dissociative electron transfer reaction $\mathrm{N}^+_2 + X \rightarrow \mathrm{N} + \mathrm{N} +X^+$ which leads to two neutral N atoms in the final state. Assuming a kinetic energy release in the order of 1~eV, both neutral fragments may reach the converter cathode within a time interval in the order of 10~ns. A fast multi-fragment detector dedicated to observation of such reactions involving multiple neutral products under CSR conditions is presently being set-up \cite{beckerdiss}. However, the electronics of the COMPACT detector is not designed to resolve such nearly-coincident double hits. In the present experiment, the two N atoms may thus appear as a single, larger anode pulse.\par 
In the photo-detachment experiments conducted at 6~K operating temperature, care was taken to keep the average rate of detected particles $\le 300$~s$^{-1}$, in order to avoid the pulse height saturation effects observed at higher count rates (cf.\ Sect.~\ref{operation} and Fig,~\ref{rates}). Generally, large values of $b$ are found also in these cases, though differences have been observed. The interplay of $\tilde{\gamma}$ and $b$ in Eq.~(\ref{polya}) has a strong impact on the coefficient $P_0$, and thereby on the detection efficiency that can maximally be achieved ($1-P_0$). A relatively small $b$ in conjunction with a high secondary yield can lead to very good detection efficiencies (as in the Co$_2^-$ experiment listed in Tab.~\ref{fitresults}, with $1-P_0 \approx 1$). The opposite case of a small yield combined with a high value of $b$ leads to the poorest detection efficiency, especially as the corresponding, nearly exponentially decreasing, pulse height distributions are particularly susceptible to the signal discrimination threshold (Ag$_2^-$).\par 
\subsection{Limits of the Efficiency Model}
The above procedure provides an independent experimental value for the detection efficiency of the COMPACT set-up that can be determined \emph{in-situ}. Although differences in the mean pulse amplitudes are often apparent without analysis, the lack of distinctive features in the pulse height distributions renders the fit model quite sensitive to statistical or systematic effects. Electronic noise picked up by the front-end amplifier can falsify the spectra to the point where reliable evaluation is not possible. Numerically, the model is very sensitive to the experimentally determined single-electron distribution $f_1$. As the latter is steeply decreasing towards higher amplitudes, the significant part of $f_1$ that lies below the discriminator threshold has to be extrapolated. Due to this, we estimate that the results from Tab.~\ref{fitresults} are bound to a systematic uncertainty of $10-20$\,\% in addition to the fit-statistical error-bars given.\par
Finally, the analysis relies on the assumption that the MCP-signals from different secondary electrons originating from the same heavy-particle impact sum up linearly. This is a safe assumption in room-temperature operation of the MCPs. At 6~K, however, the total charge that can be extracted from the channel-plates within $\sim 10$~ns might be limited, even though the individual avalanches that sum up to one anode pulse happen within different micro-channels, and even though no influence of the average particle count rate on the pulse height distribution was measured below $\sim 1000$~s$^{-1}$. While we consider this option unlikely, it is difficult to disprove. Note however that, in presence of such a lowered-gain effect in the heavy-particle data, the fit model would underestimate $\tilde{\gamma}$ and, thus, the detector efficiency, rather than overestimate them.  
\section{Summary and Outlook\label{summary}}
Using the emerging cryogenic storage rings for slow heavy ions, a set of well-established experimental techniques can be applied to a new range of low-energy molecular and atomic physics. This requires, however, that those techniques are carefully adapted to the peculiarities of that new class of storage devices. Simple, robust, and inexpensive particle detectors are still not a generally-established type of beam line instrumentation in extremely low temperature environments, although the usefulness of such devices is undisputed. Following-up on our technical design paper on the movable COMPACT detector \cite{spruck}, we have reported on the first data-taking operation of this device in the Cryogenic Storage Ring CSR \cite{csrpaper}.\par 
At a temperature of 6~K, the detector proved to operate reliably over a dynamic counting range between $3\times10^3$ and  $3\times10^5$, depending on the time structure of the particle hits. For continuous irradiation, a critical average particle count rate of the order of 1000 s$^{-1}$ has been found. The response of the detector to non-continuous daughter beams---where particles arrive in bursts rather than as a steady stream---was tested, and the device was used productively in such experiments. Systematic studies of this application will follow. The available dynamic range of the detector can likely be further improved by the built-in MCP heating, a device whose operation was successfully tested, but not yet used in the experiments reported here. We have shown how a statistical model for the pulse height distributions can be used to derive independent estimates for the detection efficiency. In various experiments, the latter was found to lie in the range $0.5 - 1.0$ also at a detector temperature of 6~K.\par
For well-localised hits of ions on the converter cathode, as anticipated, e.g., for future electron-cooled atomic parent ions, the observed secondary-electron statistics is expected to become near-Poissonian, resulting in smaller signal loss and easier pulse discrimination. However, low emittance product beams, where all particles impinge onto a defined small spot on the detector might also be more susceptible to saturation effects due to charge depletion of the MCPs at low temperature---an aspect that will need to be carefully studied. The fact that, even in this situation, the `Daly' working principle leads to a natural spread of the secondary electrons on the MCP might be another advantage of the COMPACT design.\par
On the basis of the favourable outcome of the first experiments, a second specimen of the COMPACT detector is being built. While its particle sensor and detection electronics are nearly identical to the existing detector, it features a translation stage with shorter travel range, fitting a dedicated vacuum chamber directly following the 6$^\circ$-bender in the CSR beam line (cf.\ Fig.~\ref{csrfigure}). This second detector will complement the existing COMPACT in future electron collision experiments. Due to the shorter distance to the bending element, the charge-to-mass resolution of the new set-up will be worse compared to its predecessor. However, its range of detectable product ions will be much larger, enabling it to detect, e.g., charged fragments from molecular breakup events with rigidities differing from that of the parent beam by more than 350\,\% in both directions.\par  
Finally, variants of the COMPACT set-up optimised for uncooled beams are presently being developed to be part of the experimental equipment in upcoming CSR beam-times. These follow the same principle as the original detector, but offer a much-enlarged sensitive aperture in order to intercept high-emittance product beams without geometric loss.\par 
\section{Acknowledgements}
This work would have been impossible without skilled support from the accelerator crew and mechanical workshops of the Max Planck Institute for Nuclear Physics (MPIK) which is hereby gratefully acknowledged. We thank X.~Urbain for helpful discussions on the physics of MCPs. We are grateful for the financial support obtained from the Max Planck Society (MPG). The work of A.B. and K.S. was partly funded by the German Research Foundation (DFG) under contract numbers Wo 1481/2-1 and Schi 378/9-1, respectively.   
\section*{References}
\end{document}